\documentclass[aps,prd,showpacs,groupedaddress,floatfix,nofootinbib]{revtex4}

\usepackage{graphicx}
\usepackage[symbol*]{footmisc}
\usepackage{array}
\usepackage{amsmath}
\usepackage{amssymb}
\usepackage{amsfonts}
\usepackage{verbatim} 
\usepackage{multirow} 
\usepackage{slashbox} 
\usepackage{array} 

\DeclareGraphicsExtensions{.eps, .eps, .jpg, .png}

\def\be{\begin{equation}}
\def\ee{\end{equation}}
\newcommand{\bea}{\begin{eqnarray}}
\newcommand{\eea}{\end{eqnarray}}
\newcommand{\nn}{\nonumber}
\def\ie{{\it i.e.}}

\def\d{{\rm d}}
\def\del{\partial}

\begin{document}

\title{Vacua and instantons of ghost-free massive gravity}

\author{Minjoon Park}
\email[]{minjoonp@physics.umass.edu}
\affiliation{Department of Physics, University of Massachusetts, Amherst, MA 01003}

\author{Lorenzo Sorbo}
\email[]{sorbo@physics.umass.edu}
\affiliation{Department of Physics, University of Massachusetts, Amherst, MA 01003}

\date{\today}

\begin{abstract}

Recently discovered models of ghost-free massive gravity and bigravity are characterized by a non-trivial potential that gives rise to a rich vacuum structure. We review maximally symmetric vacua of the de Rham-Gabadadze-Tolley (dRGT) massive gravity and of the Hassan-Rosen (HR) bigravity, and discuss their perturbative stability. In particular, we discuss perturbations about self-accelerating vacua in HR bigravity, and argue that, analogously to what was found in the case of dRGT gravity, some of them contain strongly coupled modes. We then show that it is impossible to construct regular instantons connecting different classically stable vacua of dRGT gravity without violating energy conservation or the null energy condition.

\end{abstract}

\pacs{04.50.Kd}

\maketitle

\section{Introduction}

A series of papers~\cite{Gabadadze:2009ja,deRham:2009rm,deRham:2010gu,deRham:2010ik,Hassan:2011hr,Hassan:2011tf,Hassan:2011ea} have recently led to the construction of models of  massive gravity (denoted as ``dRGT'' from the authors of~\cite{deRham:2010kj}) and bigravity (``HR''~\cite{Hassan:2011zd}) that propagate only five (or seven, in the case of bigravity) degrees of freedom and are therefore free of the Boulware-Deser (BD) ghost~\cite{Boulware:1973my}. Remarkably, these models are described by a finite (ranging from 3 to 6, depending on the assumptions) number of parameters. Such a parameter space is large enough to yield a rich vacuum structure, while being small enough to allow for a complete characterization of the vacua of the theory. Besides the obvious Minkowski vacua, various cosmological solutions~\cite{Gumrukcuoglu:2011ew,Volkov:2011an,vonStrauss:2011mq,Comelli:2011zm,Gratia:2012wt,Kobayashi:2012fz,Volkov:2012cf,Volkov:2012zb,Langlois:2012hk} and spherically symmetric backgrounds~\cite{Koyama:2011xz,Nieuwenhuizen:2011sq,Comelli:2011wq,Berezhiani:2011mt} have been discovered. 

In the present paper, complementing previous works, we characterize the landscape of the $SO(4)$-invariant vacua of Euclidean dRGT massive gravity and investigate their perturbative and -- to our knowledge, for the first time -- non-perturbative stability. As we will see, we will recover two disjoint branches of solutions: vacua in the form of constant factor$\,\times\,$Minkowski and (Anti-)de Sitter-like vacua. Perturbative analysis shows that depending on the choice of parameters some of these vacua will be stable, while others will be plagued by tachyonic and/or ghost-like pathologies.  

Given the presence of multiple vacua, it is natural to ask whether there can be non-perturbative instanton transitions bringing one vacuum to a different one. We will show that it is impossible to construct solutions describing such a transition unless we allow for a non-conserved source or for the violation of the null energy condition in the matter sector. These results are similar (but, as we will see, the details are quite different) to the findings of \cite{Izumi:2007gs}, who studied the possibility of non-perturbative transitions in the context of the Dvali-Gabadadze-Porrati (DGP) model~\cite{Dvali:2000hr}. Zhang {\it et al}.~\cite{Zhang:2012ap} also looked into instantons in the context of dRGT gravity. However, these authors studied instantons between the different vacua of a given scalar field, whereas in the present work we are concerned with the vacuum structure of dRGT gravity itself. It is worth pointing out that the quantum consistency of the models under consideration has not been fully explored yet, and there might be inconsistencies. For example, the structure found in the papers~\cite{deRham:2010kj,Hassan:2011zd} that guarantees the absence of the BD ghost may not be stable under radiative corrections~\cite{Park:2010rp,Buchbinder:2012wb}. Moreover, like any theory of massive gravity, these models suffer from strong coupling at low energies that would cause a loss of predictability at scales as large as a kilometer~\cite{Burrage:2012ja}. Since instanton transitions are a purely quantum effect, our effort should be understood as another probe to the quantum aspects of massive gravity.

For completeness, we will also extend our program of vacuum search to HR bigravity, finding again two classes of maximally symmetric vacua, which in the limit where the second graviton decouples converge to the corresponding vacua of dRGT. Then at perturbative level, we will study the dynamics of vector modes on one such class of vacua: self-accelerating vacua in dRGT are known to contain strongly coupled modes~\cite{Koyama:2011wx,Gumrukcuoglu:2011zh,D'Amico:2012pi,Tasinato:2012ze}. Since dRGT corresponds to HR in the limit where the second metric becomes non-dynamical, one might ask whether strong coupling is an artifact of the decoupling of the second metric. It turns out not to be the case: even when both metrics are dynamical, the self-accelerating branch contains (infinitely) strongly coupled vector modes, consistently with the findings of~\cite{Comelli:2012db}. We also study the stability of perturbations about vacua of the form of constant factor$\,\times\,$de Sitter in HR gravity. We leave the investigation of instantons in HR to future work.

The plan of the paper is the following. After introducing a general formulation of massive (bi)gravity theories in the next section, in \S\ref{sec:drgt} we will find the $SO(4)$-symmetric Euclidean vacua of dRGT, and calculate the linearized action about them to obtain criteria for perturbative stability. Then in Section \ref{sec:inst}, we check the non-perturbative stability of dRGT by trying to construct instantons for all possible configurations of vacuum transitions. Section \ref{sec:hr} is devoted to the background and the perturbative analyses of HR theory, and we conclude in \S\ref{sec:conc}. Heavy algebraic details can be found in the Appendix.

\section{Formulation of Ghost-free massive (bi)gravity}\label{sec:formulation}

The most general form of BD ghost-free massive gravity is the Hassan-Rosen (HR) bigravity~\cite{Hassan:2011zd}, where two metrics $g_{\mu\nu}$ and $f_{\mu\nu}$ are both dynamical and whose action reads
\bea\label{eqn:hract}
S = \frac{M_g^2}{2}\int\d^4x \sqrt{-g}\;{}^{(g)}R + \frac{M_f^2}{2}\int\d^4x \sqrt{-f}\;{}^{(f)}R + m^2M_{\rm eff}^2\int\d^4x \sqrt{-g}\; L_\beta + S_{\rm matter} \,,
\eea
where $M_{\rm eff}^{-2}=M_g^{-2}+M_f^{-2}$, $\gamma^\mu_\rho\gamma^\rho_\nu=g^{\mu\rho}f_{\rho\nu}$ and
\bea
L_\beta = \sum_{n=0}^4\beta_n e_{(n)}(\gamma) &=& \beta_0 + \beta_1\gamma + \frac{\beta_2}{2}(\gamma^2 - \gamma\cdot\gamma) + \frac{\beta_3}{6}(\gamma^3 - 3\gamma\, \gamma\cdot\gamma + 2\gamma\cdot\gamma\cdot\gamma) \nn\\
&&+ \frac{\beta_4}{24}\big[\gamma^4 - 6\gamma^2\gamma\cdot\gamma + 8\gamma\,\gamma\cdot\gamma\cdot\gamma + 3(\gamma\cdot\gamma)^2 - 6\gamma\cdot\gamma\cdot\gamma\cdot\gamma\,\big]\,.
\eea
In order not to reintroduce the BD ghost, matter should couple to either one of the metrics but not to both, and we assume only $g$ couples to matter. Using 
\be
\sqrt{-g} \sum_{n=0}^4\beta_n e_{(n)}(\gamma) = \sqrt{-f} \sum_{n=0}^4\beta_{4-n} e_{(n)}(\gamma^{-1}),
\ee
we get the full EOMs:
\bea\label{eqn:geom}
{}^{(g)}G^\mu_\nu + m^2\frac{M_{\rm eff}^2}{M_g^2}\Big[ \tau^\mu_\nu - \delta^\mu_\nu L_\beta \Big] &=& \frac{T^\mu_\nu}{M_g^2} \,,\\
\label{eqn:feom}
{}^{(f)}G^\mu_\nu + m^2\frac{M_{\rm eff}^2}{M_f^2}\Big[ \tilde\tau^\mu_\nu - \delta^\mu_\nu \tilde{L}_\beta \Big] &=& 0 \,,
\eea
with $T^\mu_\nu$ the stress-energy tensor for matter and
\bea
&&\tau^\mu_\nu = \sum_{n=1}^4\beta_n \varepsilon_{(n)}{}^\mu_\nu(\gamma) = \beta_1\gamma^\mu_\nu + \beta_2(\gamma\,\gamma^\mu_\nu - \gamma^\mu_\rho\gamma^\rho_\nu) + \frac{\beta_3}{2}(\gamma^2\gamma^\mu_\nu - \gamma\cdot\gamma\,\gamma^\mu_\nu - 2\gamma\,\gamma^\mu_\rho\gamma^\rho_\nu + 2\gamma^\mu_\rho\gamma^\rho_\sigma\gamma^\sigma_\nu) \nn\\
&&\hspace{100pt}+ \frac{\beta_4}{6}\big[\gamma(\gamma^2-3\gamma\cdot\gamma)\gamma^\mu_\nu + 3(\gamma\cdot\gamma-\gamma^2)\gamma^\mu_\alpha\gamma^\alpha_\nu + 2\gamma\cdot\gamma\cdot\gamma\, \gamma^\mu_\nu + 6\gamma\,\gamma^\mu_\alpha\gamma^\alpha_\beta\gamma^\beta_\nu - 6\gamma^\mu_\alpha\gamma^\alpha_\beta\gamma^\beta_\rho\gamma^\rho_\nu\,\big] \,, \nn\\
&&\tilde\tau^\mu_\nu = \sum_{n=1}^4\beta_{4-n} \varepsilon_{(n)}{}^\mu_\nu(\gamma^{-1})\,,\quad\tilde{L}_\beta = \sum_{n=0}^4\beta_{4-n} e_{(n)}(\gamma^{-1}) \,.
\eea

To obtain the dRGT model, we first decouple the reference metric $f_{\mu\nu}$ by taking $M_f\to\infty$, which also results in $M_{\rm eff}\to M_g$. In this regime the term in $\beta_4$ does not contribute to EOMs and (\ref{eqn:feom}) is solved by any Ricci-flat $f_{\mu\nu}$. For generic values of $\beta_0,\,\dots,\,\beta_3$, however, the equations of motion do not allow for a Minkowski solution $g_{\mu\nu}=\eta_{\mu\nu}$ unless $\beta_0+3\,\beta_1+3\,\beta_2+\beta_3=0$. In this case, it is customary to redefine the parameters as
\bea\label{eqn:drgtalphas}
&&\beta_0=6-4\,\alpha_3+\alpha_4\,,\quad \beta_1=-3+3\,\alpha_3-\alpha_4\,,\quad\beta_2=1-2\,\alpha_3+\alpha_4\,,\quad \beta_3=\alpha_3-\alpha_4 \,.
\eea 
Thus the dRGT model is defined with three parameters, $\alpha_3$, $\alpha_4$ and $m$, where $\alpha$'s give various interactions between $f$ and $g$, and $m$ is a parameter with the dimension of mass that will turn out to give the mass of the graviton about the vacuum with $g_{\mu\nu}=\eta_{\mu\nu}$.

\section{dRGT gravity}\label{sec:drgt}

\subsection{Vacua}\label{sec:drgtvac}

We study the vacua of the dRGT theory in Euclidean signature and we confine ourselves to $SO(4)$-symmetric metrics. Our Ansatz is then 
\bea\label{gmn}
g_{\mu\nu}\d x^\mu \d x^\nu &=& a(r)^2\,\d r^2+b(r)^2\,\d\Omega_{\rm III}^2\,, \\
\label{fmn}
f_{\mu\nu}\d x^\mu \d x^\nu &=& \d r^2+r^2\,\d\Omega_{\rm III}^2\,,
\eea
where we assume $a,b>0$, and $\d\Omega_{\rm III}^2=\d\chi^2+\sin^2\chi\,\d\Omega_{\rm II}^2=\d\chi^2+\sin^2\chi(\d\theta^2+\sin^2\theta\,\d\varphi^2)$ is the metric on $S^3$. Using the Ansatz~(\ref{gmn}-\ref{fmn}), the equation of motion (\ref{eqn:geom}) has only two independent components. But one of them being redundant, we have only
\be\label{eqn:drgteom1}
3\,b\,b'{}^2 - 3\,b\,a^2- m^2\,\big[(6-4\,\alpha_3+\alpha_4)\,b^3 - 3\,(3-3\,\alpha_3+\alpha_4)\,r\,b^2 + 3\,(1-2\,\alpha_3+\alpha_4)\,r^2\,b + (\alpha_3-\alpha_4)\,r^3\big]\,a^2=0 \,.
\ee
The other equation comes from the Bianchi identity, \ie, the covariant divergence of (\ref{eqn:geom}) should vanish:
\be\label{eqn:drgteom2}
\big(a-b'\big)\, \big[(3-3\,\alpha_3+\alpha_4)\,b^2 - 2\,(1-2\,\alpha_3+\alpha_4)\,r\,b - (\alpha_3-\alpha_4)\,r^2\,\big]=0 \,.\\
\ee
Eq.~(\ref{eqn:drgteom2}) can be solved by either $a=b'$ or $b=\chi_\pm r$, where
\be
\chi_\pm=\frac{1-2\alpha_3+\alpha_4\pm\sqrt{1-\alpha_3+\alpha_3^2-\alpha_4}}{3-3\alpha_3+\alpha_4} \,,
\ee
and we have the following two branches of solutions:
\begin{enumerate}
\item Cosmological Solution (CS): For $b=\chi_\pm\, r$, solving (\ref{eqn:drgteom1}) for $a$ gives 
\be
a^2=\frac{\chi_\pm^2}{1-\mu(\chi_\pm)\,m^2\,r^2} \,,
\ee
with 
\be
\mu(x) = \frac{1-x}{3\,x}\big[(6-4\,\alpha_3+\alpha_4)\,x^2 - (3-5\,\alpha_3+2\,\alpha_4)\,x - \alpha_3+\alpha_4\big] \,,
\ee
and the physical metric reads
\be\label{confmink}
g_{\mu\nu}\d x^\mu \d x^\nu=\frac{\chi_\pm^2}{1-\mu(\chi_\pm)\,m^2\,r^2}\d r^2+\chi_\pm^2\,r^2\,\d\Omega_{\rm III}^2 \,.
\ee
With a double Wick rotation we write the metric in Lorentzian signature as
\be
g_{\mu\nu}\d x^\mu \d x^\nu=-\frac{\chi_\pm^2}{1+\mu_\pm m^2\,\tau^2}\d\tau^2+\chi_\pm^2\tau^2(\d R^2+\sinh^2R\,\d\Omega_{\rm II}^2)\,,
\ee
with $\mu_\pm\equiv\mu(\chi_\pm)$. Since $\mu_+>0$, we can bring the metric with $\chi_+$ to a cosmological form through a redefinition $\tau=\frac{1}{m\sqrt{\mu_+}}\sinh\frac{m\sqrt{\mu_+}}{\chi_+}t$ :
\be
g_{\mu\nu}\d x^\mu \d x^\nu = -\d t^2 + \Big(\frac{\chi_+}{m\sqrt{\mu_+}}\sinh\frac{m\sqrt{\mu_+}}{\chi_+}t\Big)^2 (\d R^2+\sinh^2R\,\d\Omega_{\rm II}^2)\,. \nn\\
\ee
This is the de Sitter metric in open chart found in~\cite{Gumrukcuoglu:2011ew}. As for the solution with $\chi_-$, since $\mu_-<0$, it turns out to describe an Anti-de Sitter metric.\footnote{Arbitrary cosmological solutions were found in \cite{Gratia:2012wt,Kobayashi:2012fz}, where $f_{\mu\nu}$ took complicated forms. The de Sitter solutions  found in those papers would agree with the open solution of \cite{Gumrukcuoglu:2011ew} up to coordinate transformations. For example, after a standard coordinate transformation from open dS to flat dS, the vacuum solution of \cite{Gumrukcuoglu:2011ew} can be matched to the de Sitter solution of~\cite{Gratia:2012wt} with $f(t,r)=\frac{x_0}{2}Hr^2e^{H t}+\frac{x_0}{H}\sinh Ht$, where $H^2\equiv\frac{m^2}{6}P_0(x_0)$, and $P_0$, $f$ and $x_0$ are defined in \S III of \cite{Gratia:2012wt}.}
\item Conformal-to-Minkowski (CM): If $a=b'$, $g_{\mu\nu}$ is flat, and eq.~(\ref{eqn:drgteom1}) has three solutions: $b=r$ and $b=c_\pm r$, with
\be
c_\pm=\frac{3-5\,\alpha_3+2\alpha_4\pm\sqrt{9-6\alpha_3+9\alpha_3^2-12\alpha_4}}{2(6-4\alpha_3+\alpha_4)} \,.
\ee
Therefore we have a Minkowski and two conformal-to-Minkowski vacua:
\be
g_{\mu\nu}\d x^\mu \d x^\nu=c_i^2\,(\d r^2+r^2\,\d\Omega_{\rm III}^2) \,,
\ee
where $i=+,0,-$ and $c_0=1$. For all three solutions, the physical metric $g_{\mu\nu}$ can be transformed into Minkowski with a simple coordinate rescaling. But of course such a rescaling causes the fiducial metric $f_{\mu\nu}$ to move away from Minkowski, so that each solution corresponds to a different physical situation; while ordinary matter will behave exactly the same on the three backgrounds with $b=c_+\,r$, $b=c_-\,r$, and $b=r$, graviton fluctuations will feel a different mass. For example, as we will see below, when $6-4\,\alpha_3+\alpha_4>0$, perturbations about the metrics with $c_0$ or $c_-$ describe a graviton with positive mass, whereas perturbations about $c_+$-vacua are tachyonic. 
\end{enumerate}

\subsection{Perturbations of the CM backgrounds}

Let us now perform a perturbative analysis on top of CM backgrounds described above and investigate their stability.\footnote{Perturbations in the CS branch were investigated in \cite{Gumrukcuoglu:2011zh,DeFelice:2012mx,D'Amico:2012pi,Wyman:2012iw}.} Since only $g$ is dynamical, we expand (\ref{eqn:hract}) as $g_{\mu\nu} = c_i^2\,\eta_{\mu\nu} + h_{\mu\nu}$, where $x^\mu=(t,\vec x)$ and 
\be\label{eqn:gheldcmp}
h_{00} = -2\,\phi\,,\quad
h_{0i} = h_{0i}^{\rm T} + \del_i B\,, \quad h_{ij} = h_{ij}^{\rm TT} + \del_i\xi_j^{\rm T} + \del_j\xi_i^{\rm T} -2\,\delta_{ij}\,\psi + 2\,\del_i\del_j E \,, 
\ee
with $\del_i h_{0i}^{\rm T}=0$, $\delta_{ij}h_{ij}^{\rm TT}=0$, $\del_i h_{ij}^{\rm TT}=0$ and $\del_i\xi_i^{\rm T}=0$. To handle the expansion of $\gamma=\sqrt{g^{-1}f}$,  we use the trick of~\cite{Gumrukcuoglu:2011zh}: For $N\times N$ matrix ${\bf A}$, with
\be
{\bf A} = {\rm diag}(a_1,\cdots,a_N) + \epsilon {\bf A}^{(1)} + \epsilon^2 {\bf A}^{(2)} + {\cal O}(\epsilon^3)\,,\quad a_i>0\,,
\ee
its square-root matrix 
\be
\sqrt{\bf A} = {\rm diag}(\sqrt{a_1},\cdots,\sqrt{a_N}) + \epsilon {\bf B}^{(1)} + \epsilon^2 {\bf B}^{(2)} + {\cal O}(\epsilon^3)\,,
\ee
is given by 
\be
{\bf B}^{(1)}{}^i_j = \frac{{\bf A}^{(1)}{}^i_j}{\sqrt{a_i}+\sqrt{a_j}}\,, \quad
{\bf B}^{(2)}{}^i_j = \frac{({\bf A}^{(2)}-{\bf B}^{(1)}\cdot{\bf B}^{(1)})^i_j}{\sqrt{a_i}+\sqrt{a_j}}\,.
\ee

Since different helicities do not mix at the quadratic level of the expansion of (\ref{eqn:hract}), we consider each helicity mode separately. The quadratic action for helicity-2 modes is easy to obtain, whereas for helicity-1 and -0 modes, we have to integrate out/solve for non-dynamical modes (equivalently, one can take $M_f\to\infty$ limit of the derivation of the quadratic action of HR bigravity presented in the Appendix): 
\bea\label{eqn:hel2}
S_{\rm hel-2} &=& \frac{M_g^2}{2}\int\d x^4 \Big\{\frac{1}{4}(\dot h^{\rm TT}_{ij})^2 + \frac{1}{4}h^{\rm TT}_{ij}(\Delta-m_{\rm eff}^2) h^{\rm TT}_{ij}  \Big\}\,, \\
\label{eqn:hel1}
S_{\rm hel-1} &=& \frac{M_g^2}{2}\int\d x^4 \Big\{\frac{1}{2}\frac{m_{\rm eff}^2}{m_{\rm eff}^2-\Delta}(\sqrt{-\Delta}\,\dot \xi^{\rm T}_i)^2- \frac{1}{2}m_{\rm eff}^2(\sqrt{-\Delta}\,\xi^{\rm T}_i)^2\Big\}\,, \\
\label{eqn:hel0}
S_{\rm hel-0} &=& \frac{M_g^2}{2}\int\d x^4 \Big\{6\,\dot\psi^2 + 6\,\psi(\Delta-m_{\rm eff}^2)\,\psi\Big\}\,,
\eea
where $\Delta=\partial_i^2$ and where
\be
m_{\rm eff}^2(c_i) = m^2\Big(\frac{1-2\alpha_3+\alpha_4}{c_i^2} - \frac{2(3-3\alpha_3+\alpha_4)}{c_i}+ 6-4\alpha_3+\alpha_4\Big)\,.
\ee
It is now obvious that the sign of $m_{\rm eff}^2$ determines the stability of the model. By  plotting $m_{\rm eff}^2$ for $\alpha_3$ and $\alpha_4$, one can see that
\begin{itemize}

\item $m_{\rm eff}^2(1)=1$, so that a theory with a Minkowski vacuum is well behaved for any value of $\alpha_3$ and $\alpha_4$, at least at the linear level.

\item $m_{\rm eff}^2(c_+)$ is negative, and therefore in a model built on the $c_+$-vacuum, all the modes, (\ref{eqn:hel2}-\ref{eqn:hel0}), are tachyonic. Furthermore, the helicity-1 mode becomes a ghost at short enough scales.

\item When $6-4\,\alpha_3+\alpha_4<0$, even $m_{\rm eff}^2(c_-)$ becomes negative, so that $c_-$-vacuum as well as $c_+$-one has tachyonic instabilities.

\end{itemize}

These results can also be explained by the following argument: If we look for metrics conformal to Minkowski, $g_{\mu\nu}=\Phi^2\,\eta_{\mu\nu}$, we see that $\Phi$ has a quartic potential 
\be\label{eqn:pot}
V(\Phi) \propto \Phi\,\big[(6-4\,\alpha_3+\alpha_4)\,\Phi^3 - 4\,(3-3\,\alpha_3+\alpha_4)\,\Phi^2 + 6\,(1-2\,\alpha_3+\alpha_4)\,\Phi + 4\,(\alpha_3-\alpha_4)\big]\,,
\ee
whose extrema are located at $\Phi=c_+$, $1$ and $c_-$. For $6-4\,\alpha_3+\alpha_4>0$, it can easily be seen that $c_- < c_+ < 1$, so that $\Phi=1$ and $c_-$ are local minima, while $\Phi=c_+$ is a local maximum and therefore unstable. On the other hand, when $6-4\,\alpha_3+\alpha_4<0$, the ordering among CM vacuum solutions changes into $c_+ < 1 < c_-$. Combining this with the fact that the potential (\ref{eqn:pot}) is now unbounded from below, we see that the Minkowski vacuum is the local minimum and $c_\pm$-vacua are local maxima, hence the appearance of instabilities for both $c_+$ and $c_-$ backgrounds.

\begin{figure}
  \begin{center}
    \resizebox{.5\textwidth}{!}{\includegraphics{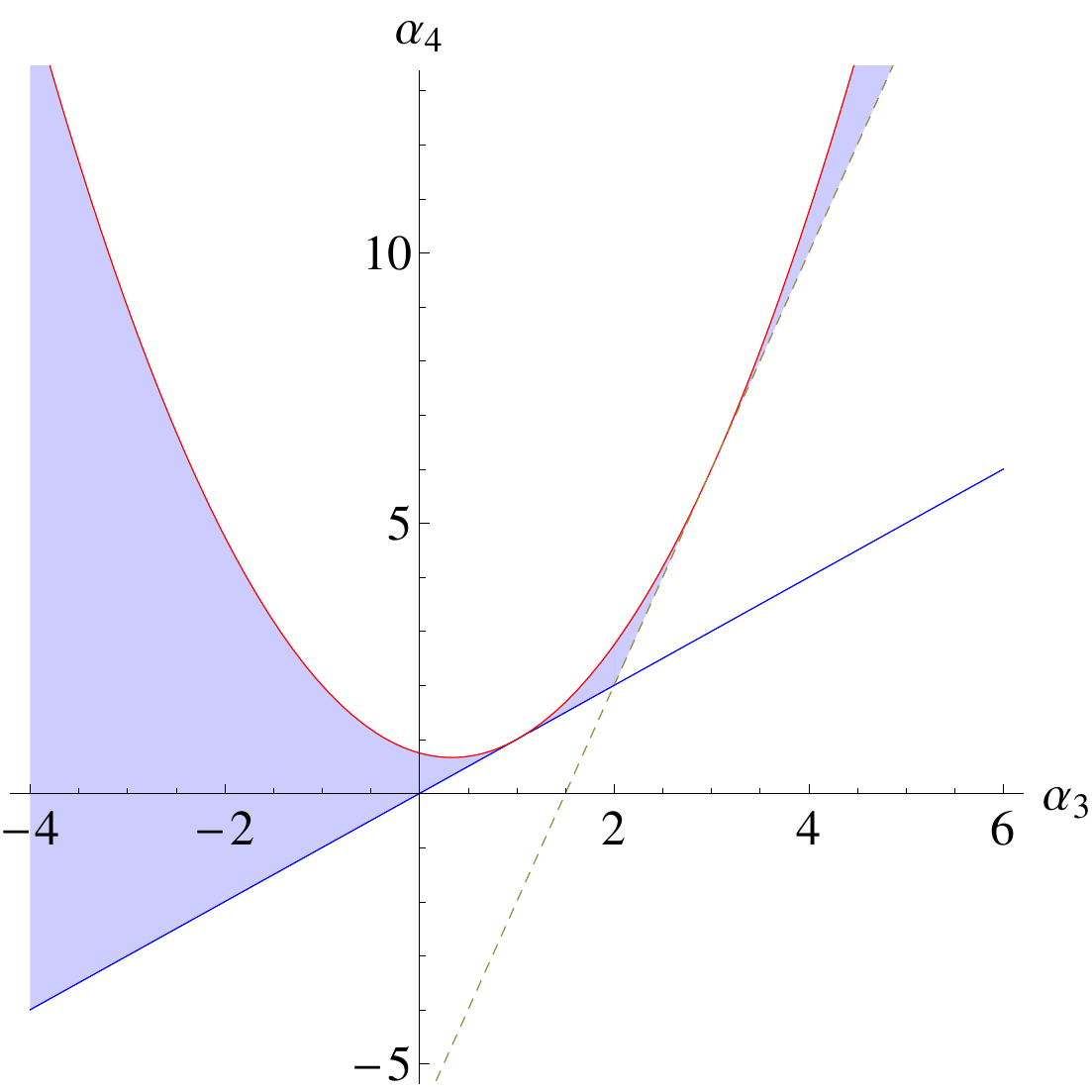}}
  \end{center}
  \caption{Marked by blue shade is the region in the $\alpha$-plane where {\em two} stable CM vacua are allowed. The two straight lines are tangent to the parabola.}
  \label{allowed}
\end{figure}

From the perturbative analysis, we may conclude that if we want dRGT to allow for two stable CM backgrounds ($c_-$ as well as $c_0$), we should constrain the parameter space of $(\alpha_3,\alpha_4)$, such that 
\be\label{eqn:allowedalpha1}
6-4\,\alpha_3+\alpha_4>0\,.
\ee
We also assumed $b>0$ at the beginning, which requires $c_\pm$ be real and positive. Under the restriction~(\ref{eqn:allowedalpha1}), this implies
\be\label{eqn:allowedalpha2}
3-2\,\alpha_3+3\,\alpha_3^2-4\,\alpha_4>0 \quad{\rm and}\quad \alpha_4-\alpha_3>0 \,.
\ee
Therefore, while Minkowski is always a stable solution, the second, stable, conformal-to-Minkowski vacuum is allowed only on the intersection of (\ref{eqn:allowedalpha1}) and (\ref{eqn:allowedalpha2}), which is depicted in Fig.~\ref{allowed}.

\section{Instantons in dRGT}\label{sec:inst}

Since dRGT gravity can have multiple classically stable vacua, it is natural to ask whether non-perturbative solutions of the equations of motion can allow for transitions between them. As long as we limit ourselves to the $SO(4)$ symmetric solutions of pure massive gravity without matter source, it is clear that no such transitions can occur -- the metrics found in \S\ref{sec:drgtvac} include all possible solutions for such a system. However, matter might be able to act as a catalyst for vacuum transitions. In this section we will first discuss the possibility that some singular form of matter supports the instanton. We will also show that, as long as matter satisfies the null energy condition and the energy momentum tensor is conserved, there are no smooth solutions describing transitions between the various vacua described above. Given that we are considering vacua belonging to separate branches, and that we will be discussing instantons supported either by singular or by regular matter source, this section is divided into several subsections.

\subsection{(No) Singular instantons within the CM branch}

A singular metric interpolating between two vacua in the CM branch should be written in a form that allows the equation of motion~(\ref{eqn:geom}) to make sense at least in a distributional form. As a consequence, the presence of the Einstein tensor built out of $g_{\mu\nu}$ requires this metric to be continuous. Since we are considering transitions within the CM branch, we must therefore choose the gauge where $g_{\mu\nu}$ is exactly Minkowski both ``inside'' and ``outside'' the instanton, whereas we allow the auxiliary metric $f_{\mu\nu}$ to have some form of discontinuity across the instanton. As a consequence, we might look for solutions of the form
\begin{eqnarray}
&&g_{\mu\nu}=\d r^2+r^2\,\d\Omega_{\rm III}^2\,,\nonumber\\
&&f_{\mu\nu}=\left\{
\begin{array}{ll}
c_-^{-2}\,\left(\d r^2+r^2\,\d\Omega_{\rm III}^2\right)\,,& r<r_0\\
\d r^2+r^2\,\d\Omega_{\rm III}^2\,,& r>r_0
\end{array}\right.\,.\label{eqn:fdisc}
\end{eqnarray}
(We are not interested in any transition involving the $c_+$-vacuum, because it is classically unstable.) However, the metric $f_{\mu\nu}$ in eq.~(\ref{eqn:fdisc}) carries curvature on the singular surface $r=r_0$ and does not correspond to dRGT gravity that requires $f_{\mu\nu}$ to be identically Ricci-flat. Then, one can try $f_{\mu\nu}\d x^\mu \d x^\nu = y'(r)^2\,\d r^2+y(r)^2\,\d\Omega_{\rm III}^2$ with $y(r)=c_-^{-1}\,r+(1-c_1^{-1})\,r\,\Theta(r-r_0)$. Such $f_{\mu\nu}$ is flat everywhere by construction, but it contains a singularity $\propto\delta(r-r_0)^2$ in $f_{rr}$, inducing a term proportional to $\delta(r-r_0)$ in $\gamma^\mu_\nu$. While this by itself is not a problem, since the theory contains powers of $\gamma^\mu_\nu$, there will be terms with powers of the Dirac delta, which are not meaningful in a distributional sense. Equivalently, one might go back to the St\"uckelberg formulation of theory by introducing a set of four scalar fields $\phi^a$, such that $f_{\mu\nu}=\delta_{ab}\,\partial_\mu\phi^a\,\partial_\nu\phi^b$, and
\begin{eqnarray}
&&\phi^1=\left\{c_-^{-1}+(1-c_1^{-1})\,\Theta(r-r_0)\right\}\,r\,\cos\chi\,,\nonumber\\
&&\phi^2=\left\{c_-^{-1}+(1-c_1^{-1})\,\Theta(r-r_0)\right\}\,r\,\cos\theta\,\sin\chi\,,\nonumber\\
&&\phi^3=\left\{c_-^{-1}+(1-c_1^{-1})\,\Theta(r-r_0)\right\}\,r\,\sin\theta\,\sin\varphi\,\sin\chi\,,\nonumber\\
&&\phi^4=\left\{c_-^{-1}+(1-c_1^{-1})\,\Theta(r-r_0)\right\}\,r\,\sin\theta\,\cos\varphi\,\sin\chi\,\nonumber\,,
\end{eqnarray}
where $\chi$, $\theta$ and $\varphi$ are the angles on $S^3$. Again, since functions $\phi^a$ are discontinuous, powers of the Dirac delta function will appear in the equations of motion~(\ref{eqn:geom}), making them  ill-defined from a distributional point of view.

Therefore we conclude that the theory does not allow for singular and yet meaningful in a distributional sense, non-perturbative solutions describing transitions between different CM vacua.

\subsection{(No) Singular or regular instantons within the CS branch}\label{sec:instdb}

In the CS branch a given choice of $\alpha_3$ and $\alpha_4$ may allow two different vacua; a de Sitter vacuum with $b=\chi_+\,r$ and an Anti-de Sitter one with $b=\chi_-\,r$. An instanton connecting $\chi_+$-vacuum to $\chi_-$-one looks like
\be\label{confmink}
g_{\mu\nu}\d x^\mu \d x^\nu=a_{\rm In}(r)^2\d r^2+\chi_{\rm In}(r)^2\,r^2\,\d\Omega_{\rm III}^2 \,,
\ee
where $a_{\rm In}$ jumps from $\frac{\chi_+}{\sqrt{1-\mu_+\,m^2\,r^2}}$ to $\frac{\chi_-}{\sqrt{1-\mu_-\,m^2\,r^2}}$, while $\chi_{\rm In}$ goes from $\chi_+$ to $\chi_-$. Now, as in the previous case, we have to choose a coordinate system where the induced metric at the junction of the two geometries is continuous, but in doing so we generate a singularity in the $f_{\mu\nu}$ metric that makes the equations of motion ill-defined from a distributional point of view.

We can go further, and prove that no regular transition, supported by matter that obeys energy conservation, can happen within the CS branch: the Bianchi identity (\ref{eqn:drgteom2}) requires $b$ should always be either $\chi_+ r$ or $\chi_- r$, \ie, $\chi_{\rm In}$ be either $\chi_+$ or $\chi_-$. If we require a smooth metric, then $\chi_{\rm In}$ cannot change at all. Note that the form of the matter does not play any role (as long as the stress-energy tensor is conserved) in this argument, and there can be no singular or regular instanton connecting different vacua within CS branch. 

\subsection{(No) Singular transitions between vacua of different branches}%

For the case of a transition between a vacuum in the CM and one in the CS branch, we can use an argument identical to that of section~\ref{sec:instdb} above. Let us consider for instance the case where we are in the CM branch,
\begin{equation}
g_{\mu\nu}\,\d x^\mu \d x^\nu=c_i^2\,(\d r_i^2+r_i^2\,\d\Omega_{\rm III}^2)\,,\qquad f_{\mu\nu}\,\d x^\mu \d x^\nu=\d r_i^2+r_i^2\,\d\Omega_{\rm III}^2\,,\qquad r_i<\bar{r}_i\,,
\end{equation}
at small radii ($r_i$ being a radial coordinate covering the ``interior'' of the instanton, and $c_i$ denoting one between $c_0(=1)$ and $c_-$), whereas at large $r$ we are in the CS branch, 
\begin{equation}
g_{\mu\nu}\,\d x^\mu \d x^\nu=\frac{\chi_n^2}{1-\mu_n\,m^2\,r_o^2}\,\d r_o^2+\chi_n^2\,r_o^2\,\d\Omega_{\rm III}^2\,,\qquad f_{\mu\nu}\,\d x^\mu \d x^\nu=\d r_o^2+r_o^2\,\d\Omega_{\rm III}^2\,,\qquad r_o>\bar{r}_o\,,
\end{equation}
where $r_o$ covers the ``exterior'' of the instanton geometry, and $\chi_n$ is either $\chi_+$ or $\chi_-$.

Then by requiring the metric $g_{\mu\nu}$ to be continuous at the junction of the two geometries we must impose $c_i\,\bar{r}_i=\chi_n\,\bar{r}_o$, which, however, implies a discontinuity $\bar{r}_i^2\left(1-c_i^2/\chi_n^2\right)$ in the coefficient of the $\d\Omega_{\rm III}^2$ term in the non-dynamical metric $f_{\mu\nu}$. It is easy to see (see also the discussion in the next subsection) that in general $c_i\neq \chi_n$. Therefore we are forced to introduce a discontinuity in $f_{\mu\nu}$ that makes the equations of motion ill-defined in a  distributional sense, and hence singular transitions between vacua of different branches are not allowed.

\subsection{(No) Regular vacuum decay within CM branch}\label{sec:cminst}

Since the various vacua of the CM branch all belong to the same solution, $a=b'$, of the constraint~(\ref{eqn:drgteom2}), it is possible to transition among them without violating energy conservation. Therefore we have to analyze in detail the dynamical equations to see what conditions must be satisfied by the matter supporting the instanton.

An Ansatz for an instanton within the CM branch can be written as
\bea\label{eqn:igmn}
g_{\mu\nu}\d x^\mu \d x^\nu &=& b_{\mathrm {In}}'(r)^2\,\d r^2+b_{\mathrm {In}}(r)^2\,\d\Omega_{\rm III}^2\,, \\
\label{eqn:ifmn}
f_{\mu\nu}\d x^\mu \d x^\nu &=& \d r^2+r^2\,\d\Omega_{\rm III}^2\,.
\eea
Then (\ref{eqn:geom}) gives
\bea
&&\frac{T^r_r}{M_g^2}= -\frac{\rho}{M_g^2} = \frac{m^2}{b_{\mathrm {In}}^3}\,(r-b_{\mathrm {In}})\,\big[(6-4\,\alpha_3+\alpha_4)\,b_{\mathrm {In}}^2- (3-5\,\alpha_3+2\,\alpha_4)\,r\,b_{\mathrm {In}} - (\alpha_3-\alpha_4)\,r^2\big] \,,\\
&&\frac{T^\chi_\chi}{M_g^2}= \frac{p}{M_g^2}= \frac{m^2}{b_{\mathrm {In}}^2\,b_{\mathrm {In}}'}\Big\{\big[3-3\,\alpha_3+\alpha_4-(6-4\,\alpha_3+\alpha_4)\,b_{\mathrm {In}}'\big]\,b_{\mathrm {In}}^2 \nn\\
&&\hspace{75pt}- 2\,\big[1-2\,\alpha_3+\alpha_4-(3-3\,\alpha_3+\alpha_4)\,b_{\mathrm {In}}'\big]\,r\,b_{\mathrm {In}} - \big[\alpha_3-\alpha_4+(1-2\,\alpha_3+\alpha_4)\,b_{\mathrm {In}}'\big]\,r^2\Big\} \,,
\eea
where $\rho$ and $p$ are the energy density and the pressure of the matter supporting the instanton. Defining $c(r)\equiv b_{\mathrm {In}}(r)/r$, we get
\be\label{eqn:nec}
-\frac{\rho+p}{M_g^2} = \frac{m^2\,r\,c'}{c^3\, b_{\mathrm {In}}'}\,\big[(3-3\,\alpha_3+\alpha_4)\,c^2 - 2\,(1-2\,\alpha_3+\alpha_4)\,c - \alpha_3+\alpha_4\big] \,,
\ee
where the left hand side has to be positive in order to satisfy the null energy condition.\footnote{For a canonically normalized scalar field $\phi$ with potential $V(\phi)$, we would have $\rho\equiv -\phi'{}^2/2\,b_{\mathrm {In}}'{}^2+V(\phi)$ and $p\equiv -\phi'{}^2/2\,b_{\mathrm {In}}'{}^2-V(\phi)$, so that the LHS of (\ref{eqn:nec}) equals $\phi'{}^2/M_g^2b_{\mathrm {In}}'{}^2\ge 0$.} Note that the zeros of $s(c)\equiv (3-3\,\alpha_3+\alpha_4)\,c^2 - 2\,(1-2\,\alpha_3+\alpha_4)\,c - \alpha_3+\alpha_4$ are at $c=\chi_\pm$, and one can easily show that within the allowed region, (\ref{eqn:allowedalpha1}-\ref{eqn:allowedalpha2}), of $\alpha$'s, 
\be\label{eqn:ordering}
c_- < \chi_- < c_+ < \chi_+ < 1\,,
\ee
\ie,
\be
s(c) \left\{ \begin{array} {ll} 
>0\,,\quad & {\mathrm {for}}\; c<\chi_-\;{\mathrm {or}}\;\;c>\chi_+\,, \\
<0\,,\quad & {\mathrm {for}}\; \chi_-<c<\chi_+\,.
\end{array} \right.
\ee
We now have all the necessary elements to show that it is impossible to find a healthy solution where $c(r)$ interpolates between $c_-$ and $1$ as $r$ ranges from $0$ to $\infty$. 

Let us first assume that $c(r\to 0)\to 1$. Then since $s(1)=1$, it follows that $c'(r\to 0)\ge 0$, \ie, if $c$ starts at $1$, it can only increase. Therefore $c$ will never be able to decrease to reach $c_-<1$. This excludes the possibility that the interior of the instanton is the vacuum with $c=1$.

When $c(r\to 0)\to c_-$, in order to reach $c=1$, $c(r)$ needs to cross first $\chi_-$ and then $\chi_+$. If we do not want to violate the null energy condition, the quantity $c'(r)s(c)/b_{\mathrm {In}}'(r)$ should stay positive as $c$ crosses, say, $\chi_-$. Now, $c(r)$ might have wiggles, and can cross $\chi_-$ many times, but let us define $r_-$ by the smallest among $r$'s solving $c(r)=\chi_-$. Since, at $r=r_-$, $c(r)$ crosses $\chi_-$ from below, $c'(r_-)> 0$ and therefore $b_{\mathrm {In}}'(r_-)=c'(r_-)\,r_-+c(r_-)>0$. Then $c'/b_{\mathrm {In}}'$ is positive at least in a small neighborhood of $r=r_-$. But in the same neighborhood, $s(c)$ changes signs from positive to negative as $c$ crosses $\chi_-$, and so does $c'(r)s(c)/b_{\mathrm {In}}'(r)$, violating the null energy condition.

Therefore we see that there is no regular instanton that can connect the two stable CM vacua.

\subsection{(No) Regular transitions between vacua of different branches}%

We would like to check whether it is possible to smoothly connect a CS background,
\be
g_{\mu\nu}\d x^\mu \d x^\nu=\frac{\chi_n^2}{1-\mu_n\,m^2\,r^2}\d r^2+\chi_n^2\,r^2\,\d\Omega_{\rm III}^2 \,,\quad n=\pm\,,
\ee
at, say, small $r$, to a CM one,
\be
g_{\mu\nu}\d x^\mu \d x^\nu=c_i^2\,(\d r^2+r^2\,\d\Omega_{\rm III}^2) \,,\quad i=0,-\,,
\ee
at large $r$. Since generic CS and CM solutions in the presence of a source can be written as
\be
g_{\mu\nu}\d x^\mu \d x^\nu=a(r)^2\d r^2+\chi_n^2\,r^2\,\d\Omega_{\rm III}^2 \quad {\rm and} \quad
g_{\mu\nu}\d x^\mu \d x^\nu=b'(r)^2\d r^2+b(r)^2\,\d\Omega_{\rm III}^2 \,,
\ee
respectively, the solutions we are looking for should behave as follows:
\begin{itemize}
\item As $r$ increases from 0 to a certain $r_*$, $a(r)^2$ smoothly deforms from $\frac{\chi_n^2}{1-\mu_n m^2\,r^2}$ to $\chi_n^2$, so that at $r_*$ we reach $g_{\mu\nu}\d x^\mu \d x^\nu=\chi_n^2\d r^2+\chi_n^2\,r^2\,\d\Omega_{\rm III}^2$ which is {\em both} CS and CM. That is, at $r=r_*$, we can also say we have a CM solution with $b=\chi_n r$.
\item Then, as $r$ grows from $r_*$ to $\infty$, $b(r)$ smoothly transforms from $\chi_n r$ to $c_i r$.
\end{itemize}

With four vacua ($\chi_+$, $\chi_-$, $c_0$ and $c_-$) to decay to and from, we need to consider 8 different cases:
\begin{center}
\setlength{\extrarowheight}{1.3pt}
\setlength{\tabcolsep}{20pt}
  \begin{tabular}{| c || c | c | c | c |} \hline 
	\backslashbox{$r<r_*$}{$r>r_*$}	& $c_0$ 	& $c_-$ 		& $\chi_+$ 	& $\chi_-$ 	\\ \hline\hline
	$c_0$				&  \multicolumn{1}{r}{} 	&			& I 			& II 	\\ \cline{1-1}\cline{4-5}
	$c_-$ 				&  \multicolumn{1}{r}{} 	&  			& III 		& IV 	\\ \hline
	$\chi_+$ 			& V 			& VI 		& \multicolumn{1}{r}{} 	& \\ \cline{1-3}
	$\chi_-$ 			& VII 		& VIII 		& \multicolumn{1}{r}{}	& \\ \hline
  \end{tabular}
\end{center}
Before we investigate each of them, let us list some features of CM solutions, which will be useful for our investigation.

\begin{enumerate}

\item From the analysis of \S\ref{sec:cminst}, we saw that if $c(r\to0) \to c_0 \;(=1)$, the null energy condition requires $c'(r\to0)>0$ and $c$ monotonically increased.

\item Similarly, one can show that if $c(r\to\infty) \to c_-$, $c'(r\to\infty)$ should be positive in order to satisfy the null energy condition.

\item In \S\ref{sec:cminst}, we also observed that when trying to connect $c_-$ and $1$ (or even $\chi_+$) with such $c(r)$ that $c(r\to0)\to c_-$, the null energy condition was violated as $c$ crossed $\chi_-$.

\item Similar violation of the null energy condition will occur when $c$ crosses $\chi_+$, if we try to connect $\chi_-$ and 1 with $c(r)$ satisfying $c(r\to\infty)\to1$.

\end{enumerate}

Now it is easy to see that there is no instanton solution for I and II: For $r<r_*$, we have a CM solution with $b(r)=c(r)\,r$ and $c(r\to0)\to 1$. Then by item 1 of the list above, $c$ keeps increasing as $r$ increases, being unable to reach $\chi_\pm\;(<1)$ at $r=r_*$. Likewise, item 2 forbids any transition of type VI and VIII, because $c(r)$ would only decease from $c_-\;(<\chi_\pm)$ as $r$ deceases from $\infty$. It is also obvious that no vacuum decay of type III and VII can occur, because of item 3 and 4, respectively.

To examine IV, we first obtain that for a CS solution, $-(\rho+p) \propto \chi_n^2r\,a' + \chi_n^2a-a^3$, so that NEC demands $\chi_n^2r\,a' > a(a^2-\chi_n^2)$. That is, if $a(r_*)=\chi_n$, $a(r)$ will keep increasing (deceasing) as $r$ gets larger and larger (smaller and smaller) than $r_*$. For the case of IV to be realized, $a(r)^2$ should interpolate between $\chi_-^2$ at $r=r_*$ and $\frac{\chi_-^2}{1-\mu_-m^2\,r^2}$ at larger $r$. But since $\mu_-<0$, $\frac{\chi_-^2}{1-\mu_-m^2\,r^2}<\chi_-^2$ for any $r$, whereas $a(r)$ gets bigger than $\chi_-$ as $r$ grows, so that $a(r)$ cannot be deformed into $\frac{\chi_-^2}{1-\mu_-m^2\,r^2}$, making it impossible to have an instanton of type IV. Similarly, V cannot be realized either: $\frac{\chi_+^2}{1-\mu_+m^2\,r^2}>\chi_+^2$ because $\mu_+>0$. But as $r$ deceases from $r_*$, $a(r)$ gets smaller than $\chi_+$.

Therefore, we can conclude that there is no regular vacuum transition between vacua from different branches.

\subsection{Vacuum stability when there is only one classically stable vacuum}

The discussion in \S\ref{sec:cminst} concerns transitions between two different stable CM vacua. For certain values of the parameters $\alpha_3$ and $\alpha_4$, however, there is only one possible vacuum. This may occur either
\begin{itemize}
\item when $6-4\,\alpha_3+\alpha_4>0$ and $3-2\,\alpha_3+3\,\alpha_3^2-4\,\alpha_4<0$, or
\item when $6-4\,\alpha_3+\alpha_4<0$.
\end{itemize}
In the former case Minkowski is the only vacuum solution and is stable, so that there is nothing further to worry about. However, in the latter situation it is possible to see,  using the analogy~(\ref{eqn:pot}), that the Minkowski vacuum is only a local minimum of an unbounded-from-below potential, and we should check its non-perturbative stability. Since $c$ is constrained to be positive, a scenario of possible catastrophe is that a bubble of infinite negative energy forming inside of a Minkowski vacuum with $c(r\to0)\to+\infty$ and $c(r\to\infty)\to1$. Repeating analysis similar to \S\ref{sec:cminst}, we require $\frac{c'}{b_{\rm In}'}s(c) >0$ all the time, or with $x\equiv r^{-1}$,
\be\label{eqn:necwx}
\frac{\frac{\d\tilde c}{\d x}}{\frac{\d\tilde c}{x\d x}-\frac{\tilde c}{x^2}}s(\tilde c) > 0\,,
\ee
where $\tilde c(x) = c(r)$. Since $\tilde c(x\to0)\to1$, we can write $\tilde c = 1 + {\cal O}(x^\alpha)$, $\alpha>0$, around $x=0$, and then using $s(1)=1$, (\ref{eqn:necwx}) gives
\be
\lim_{x\to0}\frac{\d\tilde c}{\d x} < 0 \,.
\ee
That is, as $x$ increases, $\tilde c$ will keep decreasing, never being able to reach $\infty$, so that the catastrophe will not occur. Therefore, a Minkowski vacuum, even when it seems to be a local minimum of an unbounded-from-below ``potential", is non-perturbatively stable.

\section{HR bigravity}\label{sec:hr}%

We now consider the most general setup, where $f$ is reinstated as a dynamical field, by having $M_f$ finite. This also implies all five $\beta$'s are in play.\footnote{Background solutions -- in Lorentzian signature -- we consider in this section were already found in~\cite{vonStrauss:2011mq,Comelli:2011zm}, and their perturbative analysis has been performed, although in different contexts, in~\cite{Comelli:2012db,Hassan:2012wr}.}

\subsection{Vacua}

As before, we work with the most general $SO(4)$-symmetric metric Ansatz in the Euclidean signature,
\bea\label{eqn:gmn2}
\bar g_{\mu\nu}\d x^\mu \d x^\nu &=& a_g(r)^2\,\d r^2+b_g(r)^2\,\d\Omega_{\rm III}^2\,, \\
\label{eqn:fmn2}
\bar f_{\mu\nu}\d x^\mu \d x^\nu &=& a_f(r)^2\,\d r^2+b_f(r)^2\,\d\Omega_{\rm III}^2\,,
\eea
and use gauge freedom to fix $a_f=1$. Each of (\ref{eqn:geom}) and (\ref{eqn:feom}) gives two independent equations, and there is one conservation equation (Bianchi identity). Barring redundancy, we end up with three equations for three yet-to-be determined functions, $a_g$, $b_g$ and $b_f$:
\bea\label{eqn:hreom1}
a_g^2-(b_g')^2+\frac{m^2M_{\rm eff}^2}{3\,M_g^2}\Big(\beta_0\,b_g^2+3\beta_1\,b_g\,b_f+3\,\beta_2\,b_f^2+\beta_3\,\frac{b_f^3}{b_g}\Big)\,a_g^2&=&0 \,, \\
\label{eqn:hreom2}
1-(b_f')^2+\frac{m^2M_{\rm eff}^2}{3\,M_f^2}\Big(\beta_1\,\frac{b_g^3}{b_f}+3\,\beta_2\,b_g^2+3\,\beta_3\,b_g\,b_f+\beta_4\,b_f^2\Big)&=&0 \,, \\
\label{eqn:hreom3}
(\beta_1\,b_g^2+2\,\beta_2\,b_g\,b_f+\beta_3\,b_f^2)\,(b_g'-a_g\,b_f')&=&0 \,.
\eea
From (\ref{eqn:hreom3}), again we have two separate branches of vacua.

\subsubsection{``CS"}

When $\beta_1\,b_g^2+2\,\beta_2\,b_g\,b_f+\beta_3\,b_f^2=0$, \ie, $b_g=\bar\chi_\pm\, b_f$ with $\beta_1\,\bar\chi_\pm\hspace{-5pt}{}^2+2\,\beta_2\,\bar\chi_\pm+\beta_3=0$, (\ref{eqn:hreom1}-\ref{eqn:hreom2}) are solved by
\be\label{eqn:bgcs}
b_g=\bar\chi_\pm\, b_f\,,\quad 
b_f=\frac{\sin(\sqrt{v(\bar\chi_\pm)}\,r+\theta_1)}{\sqrt{v(\bar\chi_\pm)}}\,,\quad
a_g^2=\bar\chi_\pm\hspace{-5pt}{}^2\,\frac{1-v(\bar\chi_\pm)\,b_f^2}{1-u(\bar\chi_\pm)\,b_f^2}\,,
\ee
where $\theta_1$ is an integration constant, and where we have defined 
\bea
u(x)&=&-\frac{m^2\,M_{\rm eff}^2}{3\,M_g^2\,x}\,(\beta_0\,x^3+3\,\beta_1\,x^2+3\,\beta_2\,x+\beta_3)\,,\\
v(x)&=&-\frac{m^2\,M_{\rm eff}^2}{3\,M_f^2}\,(\beta_1\,x^3+3\,\beta_2\,x^2+3\,\beta_3\,x+\beta_4)\,.
\eea
Note that both $g$- and $f$-metrics are (A)dS, with ${}^{(g)}R^\mu_\nu=\frac{3\,u(\bar\chi_\pm)}{\bar\chi_\pm^2}\delta^\mu_\nu$ and ${}^{(f)}R^\mu_\nu=3\,v(\bar\chi_\pm)\,\delta^\mu_\nu$. We denote this branch as ``CS" as it converges to the CS branch of dRGT in the limit $M_f\to\infty$.

\subsubsection{``CM"}
When $a_g=b_g'/b_f'$, the solutions for (\ref{eqn:hreom1}-\ref{eqn:hreom2}) are
\be\label{eqn:hrcmbkg}
b_g=\bar c_i\,b_f\,,\quad 
b_f=\frac{\sin(\sqrt{v(\bar c_i)}\,r+\theta_2)}{\sqrt{v(\bar c_i)}}\,,\quad 
a_g=\bar c_i\,,
\ee
where $\theta_2$ is an integration constant and $\bar c_i$ is determined by
\be\label{eqn:hrcdef}
v(\bar c_i)-u(\bar c_i)=0\,.
\ee
In this case both metrics $g_{\mu\nu}$ and $f_{\mu\nu}$ describe spheres of radius $\bar{c}_i\,v(\bar{c}_i)^{-1/2}$ and $v(\bar{c}_i)^{-1/2}$ respectively, that is (Anti)-de Sitter spaces of different curvature (analogous metrics were discussed in~\cite{Hassan:2012wr}). We denote this branch as ``CM" because it converges to the CM branch in the limit $M_f\to\infty$. We will discuss below how the two de Sitter metrics can be simultaneously written in cosmological coordinates with flat slicing and we will study perturbations on this background.

\subsection{de Sitter ``CM" backgrounds and perturbations on top of them}\label{sec:hrcdspert}

It is easy to see how we reach dRGT from HR by taking $M_f\to\infty$ limit. But obviously HR can have a  much richer variety of vacua, including dS backgrounds in flat chart. To realize such backgrounds, we take (\ref{eqn:hrcmbkg}) and choose $\theta_2=\pi/2$. Also we constrain $\beta$'s such that $v(\bar c_i)>0$, and define 
\be\label{eqn:hdef}
v(\bar c_i)=H^2\,\bar c_i^2\,.
\ee
Then, (\ref{eqn:gmn2}-\ref{eqn:fmn2}), after a Wick rotation, becomes
\bea\label{eqn:gmncds}
\bar g_{\mu\nu}\d x^\mu \d x^\nu &=& \bar c_i^2\Big(-\d t^2+\frac{\cosh^2 H\bar c_it}{H^2\bar c_i^2}\d\Omega_{\rm III}^2\Big)\,, \\
\label{eqn:fmncds}
\bar f_{\mu\nu}\d x^\mu \d x^\nu &=& -\d t^2+\frac{\cosh^2 H\bar c_it}{H^2\bar c_i^2}\d\Omega_{\rm III}^2\,,
\eea
which are conformal-to-dS metrics in the global coordinates. 

To analyze the perturbative consistency of this background, we use the planar coordinates for dS spacetime, and rescale $t$ by $t/\bar c_i$ to obtain
\bea\label{eqn:bkgans}
\bar g_{\mu\nu} &=& {\rm diag}(-1,e^{2Ht},e^{2Ht},e^{2Ht})\,, \nn\\
\bar f_{\mu\nu} &=& \frac{1}{\bar c_i^2}{\rm diag}(-1,e^{2Ht},e^{2Ht},e^{2Ht})\,,
\eea
where $\bar c_i$ and $H$ are determined by (\ref{eqn:hrcdef}) and (\ref{eqn:hdef}), \ie,
\bea\label{eqn:bkggeom}
-\frac{3\,H^2}{m^2}-\frac{M_{\rm eff}^2}{M_g^2}\Big(\beta_0 + 3\,\frac{\beta_1}{\bar c_i} + 3\,\frac{\beta_2}{\bar c_i^2} + \frac{\beta_3}{\bar c_i^3}\Big) &=& 0  \,,\\
\label{eqn:bkgfeom}
-\frac{3\,H^2}{m^2}-\frac{M_{\rm eff}^2}{M_f^2}\Big(\beta_1\, \bar c_i + 3\,\beta_2 + 3\,\frac{\beta_3}{\bar c_i} + \frac{\beta_4}{\bar c_i^2}\Big) &=& 0 \,.
\eea
Note that, by going to conformal time $\tau=-e^{-H\,t}/H$, it is easy to see that the metric $\bar{g}_{\mu\nu}$ describes de Sitter space with curvature $H^2$, whereas $\bar{f}_{\mu\nu}$ gives de Sitter space with a different curvature $(\bar{c}_i\,H)^2$. 

We now expand the HR bigravity action, eq.~(\ref{eqn:hract}), in metric perturbations. Referring the reader to Appendix A for algebraic details, the helicity-2 action is
\be
S_{\rm hel-2} = \int\d^4 x\,\Big\{\frac{1}{2}(\dot X^{\rm TT})^2 + \frac{1}{2} X^{\rm TT} \Big(\tilde\Delta + \frac{9H^2}{4}\Big) X^{\rm TT} + \frac{1}{2}(\dot Y^{\rm TT})^2 + \frac{1}{2} Y^{\rm TT} \Big(\tilde\Delta - m_{\rm eff}^2 + \frac{9H^2}{4}\Big) Y^{\rm TT} \Big\}\,,
\ee
where $X^{\rm TT}$ and $Y^{\rm TT}$ are the linear combinations of the helicity-2 perturbations of $g$ and $f$, $\tilde\Delta=e^{-2H t}\Delta$ and where
\be\label{eqn:cdsmeff}
m_{\rm eff}^2 = -m^2\,\frac{\bar c_i^2\,M_g^2 + M_f^2}{\bar c_i^3\,(M_g^2 + M_f^2)}(\beta_1\,\bar c_i^2 + 2\,\beta_2\,\bar c_i + \beta_3)\,.
\ee
For the helicity-1 and -0 sectors, we obtain the following Hamiltonians
\bea
S_{\rm hel-1} &=&\int\d^4x\,\Big\{p^{\rm T}\dot q^{\rm T} - \frac{1}{2}(p^{\rm T})^2+ \frac{1}{2}q^{\rm T}\Big(\tilde\Delta-m_{\rm eff}^2+\frac{9H^2}{4}\Big)q^{\rm T} \Big\}\,.\\
S_{\rm hel-0} &=& \int\d^4x\,\Big\{p\, \dot q - \frac{1}{2}p^2 + \frac{1}{2}q\,\Big(\tilde\Delta-m_{\rm eff}^2+\frac{9H^2}{4}\Big)q \Big\}\,,
\eea
where all the non-physical gauge DOFs are solved away, and $q^{(\rm T)}$ and $p^{(\rm T)}$ are the surviving propagating helicity-0(-1) mode(s) and its conjugate momentum.

To have a consistent theory, we need to make sure there is no ghost or tachyon. The absence of a tachyonic mode is guaranteed if $m_{\rm eff}^2 > 0$. Inspecting the field redefinitions and canonical transformations performed for the metric perturbations, we see that under the assumption of no tachyon, the hazard of a ghost lurks only in the helicity-0 sector due to the appearance of the quantity $\sqrt{m_{\rm eff}^2-2H^2}$. That is, the helicity-0 mode becomes a ghost when $m_{\rm eff}^2 < 2\,H^2$, which is the famous Higuchi ghost~\cite{Higuchi:1986py}. Therefore, requiring
\be\label{eqn:cdspsc}
m_{\rm eff}^2 > 2\,H^2 \,,
\ee
should be enough for HR bigravity on ``CM'' backgrounds to be consistent.

\subsection{An application: Vector modes in a more general ``CS" branch}\label{sec:hrcspert}

Due to its gravity-modifying nature, it is natural to try to apply dRGT to cosmology. This, however, is not trivial, as flat cosmological solutions cannot be found for simple forms of the auxiliary metric $f_{\mu\nu}$ \cite{D'Amico:2011jj}. Simple cosmological solutions can be found by looking for open spatial sections~\cite{Gumrukcuoglu:2011ew} (see our discussion in section \ref{sec:drgtvac} where these solutions appear naturally in the CS branch of our Euclidean construction). But such open solutions come with a rather worrisome feature that the vector perturbations on top of them do not propagate at the linear order \cite{Gumrukcuoglu:2011zh}, which is potentially dangerous because it could mean that they are (infinitely) strongly coupled. Ref.~\cite{DeFelice:2012mx} confirmed this concern by showing that once the background isometry is broken, an extra ghost mode appears. Additional confirmation was given in~\cite{D'Amico:2012pi}, which showed that the vector modes might get a kinetic term at higher order expansion. One may wonder if HR bigravity on a more general cosmological background can be free of such pathology. 

To check this possibility, we perform a double Wick rotation and a coordinate transformation on the metric of (\ref{eqn:bgcs}) with $\theta_1=0$, to obtain a cosmological background in HR bigravity:
\bea\label{eqn:bgcsgbkg}
&&\bar g_{\mu\nu}\d x^\mu\d x^\nu = -\d t^2+\Big(\frac{\bar\chi_\pm}{\sqrt{u(\bar\chi_\pm)}}\sinh\frac{\sqrt{u(\bar\chi_\pm)}}{\bar\chi_\pm}t\Big)^2k^2\sigma_{ij}\d x^i\d x^j\,,\\
\label{eqn:bgcsfbkg}
&&\bar f_{\mu\nu}\d x^\mu\d x^\nu = -\frac{\cosh^2\frac{\sqrt{u(\bar\chi_\pm)}}{\bar\chi_\pm}t}{\bar\chi_\pm\hspace{-5pt}{}^2\Big(1+\frac{v(\bar\chi_\pm)}{u(\bar\chi_\pm)}\sinh^2\frac{\sqrt{u(\bar\chi_\pm)}}{\bar\chi_\pm}t\Big)}\d t^2+\Big(\frac{1}{\sqrt{u(\bar\chi_\pm)}}\sinh\frac{\sqrt{u(\bar\chi_\pm)}}{\bar\chi_\pm}t\Big)^2k^2\sigma_{ij}\d x^i\d x^j\,,
\eea
with $\sigma_{ij}=\delta_{ij}-\frac{k^2\delta_{im}\delta_{jn}x^mx^n}{1+k^2\delta_{mn}x^mx^n}$, $x^i=x,y,z$. Since we are interested in the kinetic coefficients of the vector modes, let us focus on 
\be\label{eqn:hrcsvecpert}
h_{\mu\nu} = 
\begin{pmatrix}
0 & h^{\rm T}_{0i} \\ h^{\rm T}_{j0} &  \tilde\nabla_i\xi^{\rm T}_j+\tilde\nabla_j\xi^{\rm T}_i
\end{pmatrix}, \;
\theta_{\mu\nu} = 
\begin{pmatrix}
0 & \theta^{\rm T}_{0i} \\ \theta^{\rm T}_{j0} &  \tilde\nabla_i\zeta^{\rm T}_j+\tilde\nabla_j\zeta^{\rm T}_i
\end{pmatrix},
\ee
where $h$ and $\theta$ are perturbations on top of the background defined by (\ref{eqn:bgcsgbkg}-\ref{eqn:bgcsfbkg}), $\tilde\nabla_i$ is a covariant derivative with respect to $\sigma_{ij}$, and $\tilde\nabla^i h^{\rm T}_{0i}=\tilde\nabla^i \theta^{\rm T}_{0i}=0=\tilde\nabla^i \xi^{\rm T}_i=\tilde\nabla^i \zeta^{\rm T}_i$. We can then perform a coordinate transformation to fix the gauge to $\xi^T_i=0$.

It is now straightforward to expand the equations of motion (\ref{eqn:geom}) for the metric $g_{\mu\nu}$ in terms of (\ref{eqn:hrcsvecpert}), and at ${\cal O}(h,\theta)$ the $0i$-component of (\ref{eqn:geom}) gives $h^{\rm T}_{0i}=0$. Using this result, the ${\cal O}(h,\theta)$-piece of the $ij$-component of (\ref{eqn:geom}) becomes
\be\label{eqn:gijeom}
\tilde\nabla_i\zeta^{\rm T}_j + \tilde\nabla_j\zeta^{\rm T}_i=0\,,
\ee
\ie, $\zeta^{\rm T}_i=0$.

In order to repeat the same algebra for the $f$-metric EOMs (\ref{eqn:feom}), it is convenient to perform another temporal coordinate transformation by 
\be\label{eqn:hrcsttransf}
t=\frac{\bar\chi_\pm}{\sqrt{u(\bar\chi_\pm)}}\ln\left(\sqrt{\frac{u(\bar\chi_\pm)}{v(\bar\chi_\pm)}}\sinh\sqrt{v(\bar\chi_\pm)}\tau+\sqrt{1+\frac{u(\bar\chi_\pm)}{v(\bar\chi_\pm)}\sinh^2\sqrt{v(\bar\chi_\pm)}\tau}\,\right)\,,
\ee
to get
\bea\label{eqn:bgcsgbkg2}
&&\bar g_{\mu\nu}\d x^\mu\d x^\nu = -\frac{\bar\chi_\pm\hspace{-5pt}{}^2\cosh^2\sqrt{v(\bar\chi_\pm)}\tau}{1+\frac{u(\bar\chi_\pm)}{v(\bar\chi_\pm)}\sinh^2\sqrt{v(\bar\chi_\pm)}\tau}\d \tau^2+\Big(\frac{\bar\chi_\pm}{\sqrt{v(\bar\chi_\pm)}}\sinh\sqrt{v(\bar\chi_\pm)}\tau\Big)^2k^2\sigma_{ij}\d x^i\d x^j\,,\\
\label{eqn:bgcsfbkg2}
&&\bar f_{\mu\nu}\d x^\mu\d x^\nu = -\d \tau^2+\Big(\frac{\sinh\sqrt{v(\bar\chi_\pm)}\tau}{\sqrt{v(\bar\chi_\pm)}}\Big)^2k^2\sigma_{ij}\d x^i\d x^j\,,
\eea
so that perturbed equations of motion for the $f$-metric can be easily calculated. This transformation also affects the metric perturbations, and we will denote them as $\tilde h_{\mu\nu}$ and $\tilde\theta_{\mu\nu}$. Now, since (\ref{eqn:hrcsttransf}) involves only the $t$-coordinate, $\tilde h^{\rm T}_{0i}$ and $\tilde\theta^{\rm T}_{0i}$ are different from $h^{\rm T}_{0i}$ and $\theta^{\rm T}_{0i}$ only by a factor of $\frac{\d t}{\d\tau}$, while $\tilde\xi^{\rm T}_i$ and $\tilde\zeta^{\rm T}_i$ are the same as $\xi^{\rm T}_i$ (that we have set to vanish by a spatial gauge transformation) and $\zeta^{\rm T}_i$. That is, the transformation (\ref{eqn:hrcsttransf}) does not mix up perturbations. 
Then after solving away $\tilde\theta^{\rm T}_{0i}$, the ${\cal O}(h,\theta)$-part of the $ij$-component of (\ref{eqn:feom}) turns out to be the same as (\ref{eqn:gijeom}). (To be precise, the full expression of the $ij$-component of (\ref{eqn:geom}) is (\ref{eqn:gijeom}) multiplied by a complicated function of $t$, whereas that of (\ref{eqn:feom}) is (\ref{eqn:gijeom}) multiplied by a slightly different function of $\tau$.)

This result implies that, at the linear level, the equations of motion around the ``CS'' branch do not allow for propagating vectors, precisely as was found in the corresponding branch of the dRGT model. This is consistent with the results obtained in~\cite{Comelli:2012db}. Since the dRGT vector modes obtain kinetic terms from the cubic expansion of the action, and dRGT is a special case of HR, it is straightforward to argue that the higher order expansion of the HR action will provide kinetic terms for its vector modes, so that even in HR the vector perturbations are strongly coupled on cosmological backgrounds.

\section{Conclusions}\label{sec:conc}

dRGT massive gravity has a rich and interesting vacuum structure. We have discussed the most general $SO(4)$-symmetric Euclidean vacuum solutions to the theory, recovering self-accelerating~\cite{Gumrukcuoglu:2011ew} as well Minkowski-like~\cite{Comelli:2011wq} metrics. 

After reviewing the spectrum of linearized perturbations about the Minkowski-like backgrounds (those about curved backgrounds have been studied in detail, {\em e.g.}, in \cite{Gumrukcuoglu:2011zh,DeFelice:2012mx,D'Amico:2012pi,Fasiello:2012rw}), we have considered the possibility of non-perturbative transitions connecting different vacua. Remarkably, we have found that it is not possible to construct instantons supported by singular matter distributions if we require the equations of motion to make sense at least from a distributional point of view. Indeed, the continuity of the physical metric $g_{\mu\nu}$ cannot be maintained without inducing curvature singularities in the fiducial metric $f_{\mu\nu}$, while keeping a distributionally well-defined mass term.

Then we moved on to discuss instanton transitions supported by regular matter distributions. Again, our result is negative: we have shown that there can be no smooth $SO(4)$ instantons connecting the various vacua, unless we are willing to give up energy conservation or the null energy condition. More specifically, it is not possible to transition to/from a self-accelerating branch without violating energy conservation, while the null energy condition forbids any transition involving conformal-to-Minkowski backgrounds.

The situation is different from the one discussed in~\cite{Izumi:2007gs} that analyzed instanton transitions in the DGP model. There it was shown that it is not possible to leave the self-accelerating branch without violating the null energy condition. Since the self-accelerating branch of the DGP model contains a ghost (that does violate energy conditions), it is then conceivable that a kink built out of such a ghost might provide the required negative energy~\cite{Kaloper:2011qc}. In dRGT gravity a ghost is not sufficient to depart from the self accelerating branch, since, as we have seen, one needs an even more dramatic form of pathology: energy non-conservation.

We have also discussed some aspects of the landscape of vacua in the Hassan-Rosen bigravity, constructing the full set of vacua that converge, in the limit where the auxiliary metric $f_{\mu\nu}$ becomes non-dynamical, to the CS and CM solutions of dRGT discussed in section~\ref{sec:drgtvac}. In particular, we have shown that the branch converging to the CM background has well-behaved perturbations as long as the Higuchi bound is obeyed. We have also shown -- consistently with the findings of \cite{Comelli:2012db} -- that vector modes are non-dynamical in the branch of bigravity solutions that, in the limit $M_f\to\infty$, converges to the self-accelerating CS branch. With this observation we can argue that this branch will feature strongly coupled (and possibly ghost-like) modes, in analogy with the situation discussed in \cite{Gumrukcuoglu:2011zh}.

For  what concerns future work, it would be interesting to extend the instanton analysis to HR bigravity. Also, since  it was shown in \cite{DeFelice:2012mx} that anisotropic backgrounds can generate ghost-like kinetic terms for some of the fluctuations about the CS solution, it might be possible to circumvent our no-go results of section~\ref{sec:inst} by going beyond $SO(4)$-symmetric configurations. Even more important, of course, would be to find other vacua of these models with desirable features.

\acknowledgements

We thank Claudia de Rham, Matteo Fasiello, Nemanja Kaloper, Seung-yeop Lee and Andrew Tolley for very interesting discussions. This work is partially supported by the U.S. National Science Foundation grant PHY-1205986.

\appendix
\section{Perturbative expansion of HR model on the de Sitter ``CM'' background}
We expand the HR bigravity action, (\ref{eqn:hract}), in metric perturbations,
\be
g_{\mu\nu} = \bar g_{\mu\nu} + h_{\mu\nu}\,,\quad
f_{\mu\nu} = \bar f_{\mu\nu} + \theta_{\mu\nu}\,,
\ee
where $\bar g$ and $\bar f$ are given by (\ref{eqn:bkgans}) and 
\bea\label{eqn:gheldcmp}
&& h_{00} = -2\phi\,,\quad
h_{0i} = h_{0i}^{\rm T} + e^{2H t}\del_i B\,, \quad 
h_{ij} = h_{ij}^{\rm TT} + \del_i\xi_j^{\rm T} + \del_j\xi_i^{\rm T} + e^{2H t}(-2\delta_{ij}\psi + 2\del_i\del_j E) \,, \\
\label{eqn:fheldcmp}
&& \theta_{00} = -2\varphi\,, \quad
\theta_{0i} = \theta_{0i}^{\rm T} + e^{2H t}\del_i {\cal B}\,, \quad
\theta_{ij} = \theta_{ij}^{\rm TT} + \del_i\zeta_j^{\rm T} + \del_j\zeta_i^{\rm T} + e^{2H t}(-2\delta_{ij}\omega + 2\del_i\del_j {\cal E}) \,,
\eea
with $\del_i h_{0i}^{\rm T}=\del_i \theta_{0i}^{\rm T}=0$, $\delta_{ij}h_{ij}^{\rm TT}=\delta_{ij}\theta_{ij}^{\rm TT}=0$, $\del_i h_{ij}^{\rm TT}=\del_i \theta_{ij}^{\rm TT}=0$ and $\del_i\xi_i^{\rm T}=\del_i\zeta_i^{\rm T}=0$. 

\subsection{Helicity-2}
We define the linear combinations of $h^{\rm TT}_{ij}$ and $\theta^{\rm TT}_{ij}$ by
\be
h^{\rm TT}_{ij} = \frac{2e^{H t/2}}{\sqrt{\bar c_i^2M_g^2+M_f^2}}\Big(\bar c_iX^{\rm TT} + \frac{M_f}{M_g}Y^{\rm TT}\Big)\,,\quad
\theta^{\rm TT}_{ij} = \frac{2e^{H t/2}}{\sqrt{\bar c_i^2M_g^2+M_f^2}}\Big(\frac{X^{\rm TT}}{\bar c_i} - \frac{M_g}{M_f}Y^{\rm TT}\Big)\,,
\ee
where the spatial indicies in $X$ and $Y$ are suppressed. Then $X^{\rm TT}$ and $Y^{\rm TT}$ do not mix in the helicity-2 part of (\ref{eqn:hract}), so that we obtain
\be
S_{\rm hel-2} = \int\d^4 x\,\Big\{\frac{1}{2}(\dot X^{\rm TT})^2 + \frac{1}{2} X^{\rm TT} \Big(\tilde\Delta + \frac{9H^2}{4}\Big) X^{\rm TT} + \frac{1}{2}(\dot Y^{\rm TT})^2 + \frac{1}{2} Y^{\rm TT} \Big(\tilde\Delta - m_{\rm eff}^2 + \frac{9H^2}{4}\Big) Y^{\rm TT} \Big\}\,,
\ee
where $m_{\rm eff}^2$ is defined in (\ref{eqn:cdsmeff}). Note that we have used (\ref{eqn:bkggeom}-\ref{eqn:bkgfeom}) to simplify the result.

\subsection{Helicity-1}
We define new variables similar to the helicity-2 case,
\be
h^{\rm T}_{0i},\; \xi^{\rm T}_i = \frac{\sqrt{2}}{\sqrt{\bar c_i^2M_g^2+M_f^2}}\Big(\bar c_iX^{\rm T}_{(1,2)} + \frac{M_f}{M_g}Y^{\rm T}_{(1,2)}\Big)\,,\quad
\theta^{\rm T}_{0i},\; \zeta^{\rm T}_i = \frac{\sqrt{2}}{\sqrt{\bar c_i^2M_g^2+M_f^2}}\Big(\frac{X^{\rm T}_{(1,2)}}{\bar c_i} - \frac{M_g}{M_f}Y^{\rm T}_{(1,2)}\Big)\,,
\ee
with the $i$-indicies suppressed in $X^{\rm T}_{(1,2)},Y^{\rm T}_{(1,2)}$, to split the helicity-1 part of (\ref{eqn:hract}) into $X^{\rm T}$ and $Y^{\rm T}$ sectors.

In the $X^{\rm T}$ sector, we introduce a gauge invariant variable $v^{\rm T}_{(1)}$ by $X^{\rm T}_{(1)} = e^{H t/2}v^{\rm T}_{(1)} + \dot X^{\rm T}_{(2)} - 2H X^{\rm T}_{(2)}$, to see that $X^{\rm T}_{(2)}$ completely disappears from the action. $v^{\rm T}_{(1)}$ shows up as
\be
\int\d^4x\,\Big\{-\frac{1}{2}\Delta (v^{\rm T}_{(1)})^2\Big\}\,,
\ee
\ie, $v^{\rm T}_{(1)}=0$ on shell, and there is nothing left in the $X^{\rm T}$ sector.

In the $Y^{\rm T}$ sector, we first rescale $Y^{\rm T}$'s by $Y^{\rm T}_{(1,2)}=e^{H t/2}u^{\rm T}_{(1,2)}$ to get
\bea\label{eqn:hel1act}
&&S_{\rm hel-1} = \int\d^4x\,L_{\rm hel-1} = \int\d^4x\,\Big\{-\frac{1}{2} e^{2H t}u^{\rm T}_{(1)} (\tilde\Delta-m_{\rm eff}^2) u^{\rm T}_{(1)} - \frac{1}{2} \dot u^{\rm T}_{(2)} \Delta\dot u^{\rm T}_{(2)} \nn\\
&&\hspace{140pt}- \frac{1}{2}\Big(\frac{9H^2}{4}-m_{\rm eff}^2\Big) u^{\rm T}_{(2)} \Delta u^{\rm T}_{(2)} + \dot u^{\rm T}_{(2)} \Delta u^{\rm T}_{(1)} - \frac{3}{2}H u^{\rm T}_{(1)} \Delta u^{\rm T}_{(2)}\Big\} \,.
\eea
Then we employ the Hamiltonian formulation and define the momentum $P^{\rm T}$ conjugate to $u^{\rm T}_{(2)}$ by $P^{\rm T}\equiv\frac{\delta L_{\rm hel-1}}{\delta \dot u^{\rm T}_{(2)}}$. Writing (\ref{eqn:hel1act}) in $\int\d^4x(P^{\rm T}\dot u^{\rm T}_{(2)}-{\rm Hamiltonian})$ form, integrating out the non-dynamical constraint $u^{\rm T}_{(1)}$ and performing a canonical transformation by\footnote{Here we assume $m_{\rm eff}^2>0$, \ie, no tachyonic instability. If $m_{\rm eff}^2<0$, a slightly different canonical transformation is necessary, with a prefactor of $\sqrt{-m_{\rm eff}^2+\Delta}$, which becomes imaginary at small enough scales. That is, at small enough scales the helicity-1 modes have a negative definite Hamiltonian and become ghosts, just as seen in the example of dRGT perturbative analysis.}, 
\be
P^{\rm T} = \frac{\sqrt{-\Delta}}{m_{\rm eff}}\Big\{\frac{3H}{2}p^{\rm T} + \Big(\frac{9H^2}{4}-m_{\rm eff}^2\Big)q^{\rm T}\Big\} \,, \quad
u^{\rm T}_{(2)} = \frac{(-\Delta)^{-1/2}}{m_{\rm eff}}\Big(p^{\rm T} + \frac{3H}{2}q^{\rm T}\Big) \,,
\ee
put the helicity-1 action in the canonical form:
\be
S_{\rm hel-1} = \int\d^4x\,\Big\{p^{\rm T}\dot q^{\rm T} - \frac{1}{2}(p^{\rm T})^2 + \frac{1}{2}q^{\rm T}\Big(\tilde\Delta-m_{\rm eff}^2+\frac{9H^2}{4}\Big)q^{\rm T} \Big\}\,.
\ee

\subsection{Helicity-0}
Again we combine the helicity-0 modes from the $g$-metric and those from the $f$-metric by
\bea
\phi,B,\psi,E &=& \frac{\sqrt{2}}{\sqrt{\bar c_i^2M_g^2+M_f^2}}\Big(\bar c_iX_{(1,2,3,4)} + \frac{M_f}{M_g}Y_{(1,2,3,4)}\Big)\,,\nn\\
\varphi,{\cal B},\omega,{\cal E} &=& \frac{\sqrt{2}}{\sqrt{\bar c_i^2M_g^2+M_f^2}}\Big(\frac{X_{(1,2,3,4)}}{\bar c_i} - \frac{M_g}{M_f}Y_{(1,2,3,4)}\Big)\,,
\eea
so that there is no mixing between $X$ and $Y$ modes in the helicity-0 action.

All the $X$ modes, just like the helicity-1 case, turn out to be either a gauge DOF or non-dynamical, so that none of them survives onshell. In the $Y$-sector, we repeat the process of obtaining the canonical action for the helicity-1 modes. That is, we first rescale $Y$'s by $Y_{(1,3,4)}=e^{-3H t/2}\sigma_{1,3,4}$ and $Y_{(2)}=e^{-7H t/2}\sigma_{2}$ to have 
\bea\label{eqn:hel0act}
S_{\rm hel-0} &=& \int\d^4x\,\Big\{-6H^2\sigma_1^2 - 4H\sigma_1\tilde\Delta\sigma_2 - \frac{1}{2}m_{\rm eff}^2\sigma_2\tilde\Delta\sigma_2 + \sigma_1(4\tilde\Delta-6m_{\rm eff}^2+18H^2)\sigma_3 + 6H\sigma_2\tilde\Delta\sigma_3 \nn\\
&&\qquad\quad - \sigma_3\Big(2\tilde\Delta-6m_{\rm eff}^2-\frac{27H^2}{2}\Big)\sigma_3 -2(3H^2-m_{\rm eff}^2)\sigma_1\Delta\sigma_4 + (9H^2-4m_{\rm eff}^2)\sigma_3\Delta\sigma_4 \\
&&\qquad\quad - 12H\sigma_1\dot\sigma_3 - 4\dot\sigma_3\tilde\Delta\sigma_2 - 6\dot\sigma_3^2 + 4H\dot\sigma_4\Delta\sigma_1 + 4\dot\sigma_3\Delta\dot\sigma_4\Big\} \,.
\eea
Then we introduce momenta, $P_3$ and $P_4$, conjugate to $\sigma_3$ and $\sigma_4$, and write (\ref{eqn:hel0act}) in $\int\d^4x(P_3\dot\sigma_3+P_4\dot\sigma_4-{\rm Hamiltonian})$ form:
\bea
S_{\rm hel-0} &=& \int\d^4x\,\Big[P_3\dot\sigma_3 + P_4\dot\sigma_4 + \sigma_1\big\{H P_3+(4\tilde\Delta-6m_{\rm eff}^2+18H^2)\sigma_3 - 2(3H^2-m_{\rm eff}^2)\Delta\sigma_4\big\} \nn\\
&&\qquad\quad- \frac{1}{2}m_{\rm eff}^2\sigma_2\tilde\Delta\sigma_2 +\sigma_2(-e^{-2H t}P_4+6H\tilde\Delta\sigma_3) - \frac{1}{4}P_3\Delta^{-1}P_4 - \frac{3}{8}P_4\Delta^{-2}P_4 \nn\\
&&\qquad\quad-\sigma_3\Big(2\tilde\Delta-6m_{\rm eff}^2+\frac{27H^2}{2}\Big)\sigma_3 + (9H^2-4m_{\rm eff}^2)\sigma_3\Delta\sigma_4\Big] \,.
\eea 
$\sigma_2$ is a non-dynamical constraint and can be integrated away, whereas $\sigma_1$ is a Lagrange multiplier whose elimination allows us to remove $P_3$ by $P_3 = H^{-1}\big\{-(4\tilde\Delta-6m_{\rm eff}^2+18H^2)\sigma_3 + 2(3H^2-m_{\rm eff}^2)\Delta\sigma_4\big\}$. Thus we have only one dynamical conjugate pair ($\sigma_4$, $P_4$), and with a canonical transformation by
\bea
P_4 &=& \frac{2\Delta}{H}\Big\{\frac{m_{\rm eff}\sqrt{m_{\rm eff}^2-2H^2}}{2\sqrt{3}}q - (m_{\rm eff}^2-3H^2)\sigma_3\Big\}\,, \\
\sigma_4 &=& -\frac{\sqrt{3}\Delta^{-1}}{m_{\rm eff}\sqrt{m_{\rm eff}^2-2H^2}}\Big\{H p + \Big(\frac{\tilde\Delta}{3}-\frac{m_{\rm eff}^2}{2}+\frac{3H^2}{2}\Big)q\Big\}\,,
\eea
we finally get
\be
S_{\rm hel-0} = \int\d^4x\,\Big\{p \,\dot q - \frac{1}{2}p^2 + \frac{1}{2}q\Big(\tilde\Delta-m_{\rm eff}^2+\frac{9\,H^2}{4}\Big)q \Big\}\,.
\ee

\bibliography{PS12}

\end{document}